\documentclass[aps,pre,twocolumn,superscriptaddress]{revtex4}  
\usepackage{graphicx,epsfig}  
\usepackage{dcolumn}   
\usepackage{bm}        
\usepackage{amssymb}   
\usepackage{color}
\usepackage{mathtools}
\usepackage{esvect}
\usepackage{float}
\begin{document}

\title{ Mode conversion and laser energy absorption by plasma under an inhomogeneous external magnetic field}


\author{Srimanta Maity}
\email {srimantamaity96@gmail.com}
\affiliation{Department of Physics, Indian Institute of Technology Delhi, Hauz Khas, New Delhi 110016, India}

\author{Laxman Prasad Goswami}
\affiliation{Department of Physics, Indian Institute of Technology Delhi, Hauz Khas, New Delhi 110016, India}

\author{Ayushi Vashistha}
\affiliation{Institute for Plasma Research, HBNI, Bhat, Gandhinagar 382428, India}
\affiliation{Homi Bhabha National Institute, Mumbai, 400094, India}

\author{Devshree Mandal}
\affiliation{Institute for Plasma Research, HBNI, Bhat, Gandhinagar 382428, India}
\affiliation{Homi Bhabha National Institute, Mumbai, 400094, India}

\author{Amita Das}
\email {amita@iitd.ac.in}
\affiliation{Department of Physics, Indian Institute of Technology Delhi, Hauz Khas, New Delhi 110016, India}

\begin{abstract}
 
The interaction of a high-frequency laser with plasma in the presence of an inhomogeneous external magnetic field has been studied here with the help of Particle-In-Cell simulation. It has been shown that laser enters inside the plasma as an extraordinary wave (X-wave), where the electric field of the wave oscillates perpendicular to both external magnetic field and propagation direction, and as it travels through the plasma, its dispersion property changes due to the inhomogeneity of the externally applied magnetic field. Our study shows that the X-wave's electromagnetic energy is converted to an electrostatic mode as it encounters the upper-hybrid (UH) resonance layer. In the later stage of the evolution, this electrostatic wave breaks and converts its energy to electron kinetic energy. Our study reveals two additional processes involved in decaying electrostatic mode at the UH resonance layer. We have shown that the energy of the electrostatic mode also converts to a low-frequency lower-hybrid mode and high-frequency electromagnetic harmonic radiations at the resonance layer. The dependence of energy conversion processes on the gradient of external magnetic field has also been studied and analyzed.

\end{abstract}

\maketitle

 \section{\it Introduction}
 \label{intro}

The absorption of electromagnetic (EM) energy by the plasma particles has remained as one of the most important research areas in the context of laser-plasma interaction \cite{kaw2017nonlinear}. Such research is not only of fundamental interests, but also has many applications concerning inertial confinement fusion \cite{brueckner1974laser, atzeni2004physics, zohuri2017inertial}, particle acceleration schemes \cite{tajima1979laser, joshi1984ultrahigh, modena1995electron, macchi2013ion, ratan2015, ratan2016, ratan2020}, generation of x-ray \cite{rousse2004production, corde2013femtosecond} and gamma-ray sources \cite{cipiccia2011gamma}, laboratory astrophysics \cite{remington2000review}, electromagnetic localized structures \cite{soliton4,soliton5, soliton6, soliton7, soliton8, soliton9, mandal_vashistha_das_2020} etc. In most of these applications, the conversion of EM energy of the incident laser to the kinetic energy of the plasma particles is essential. There exists many well known mechanisms e.g., electron-plasma resonance absorption \cite{manes1977polarization, pearlman1977polarization, estabrook1978properties}, vacuum heating \cite{brunel1987not}, sheath-transit absorption \cite{yang1995absorption, lefebvre1997nonlinear}, and $\vec{J}\times \vec{B}$ mechanism \cite{kruer1985j} to heat the plasma particles by an intense laser pulse. The first three processes will not work in the case of normal incidence, as in that case, there is no laser electric field component along the direction of the density gradient. The $\vec{J}\times \vec{B}$ mechanism is valid even in the case of normal incidence, but the intensity of the laser should be very high such that the Lorentz force due to the laser magnetic field can no longer be neglected. These energy absorption processes are well explored and require no external magnetic field. However, the presence of an external magnetic field introduces a novel rich variety of dynamics in the interaction process \cite{vashistha2020new, maity2021harmonic, mandal2021electromagnetic, Goswami_2021}. While the laser energy can not penetrate an unmagnetized overdense plasma, it can propagate inside the bulk region of the magnetized plasma. Thus, the presence of an external magnetic field may lead to new unexplored mechanisms related to particle heating in the context of laser-plasma interaction.

 One of the most efficient mechanisms for electron heating in a magnetized plasma is electron-cyclotron-resonance (ECR) \cite{geller2018electron}. In ECR, the magnetic field lines are parallel to the EM wave propagation direction, and electrons rotate in the same direction as the electric field of EM wave in a plane perpendicular to the direction of the external magnetic field. This technique is widely used for heating purpose in fusion plasmas \cite{bornatici1983electron, erckmann1994electron, laqua1997resonant}, plasma thruster \cite{ganguli2019evaluation}, etc. Another heating mechanism in a magnetized plasma is the mode conversion into an upper-hybrid (UH) wave \cite{stix1965radiation}, which propagates perpendicular to the external magnetic field. In this case, the resonance frequency is the upper hybrid frequency ($\omega_{uh}$), where Lorentz force is responsible for the coupling between electromagnetic and electrostatic components of the field. Thus, even in the case of normal incidence and lower intensity EM fields, UH resonance offers a very useful technique for heating the plasma electrons. There were theoretical \cite{lin1981nonlinear, lin1982plasma, lin1983computer}, as well as experimental \cite{stenzel1971upper, grek1973observation, peng1978microwave} works reported where incident microwave energy was absorbed by the plasma particles at the UH resonance layer. Lin et al., \cite{lin1981nonlinear} had reported a mode conversion from upper-hybrid to lower-hybrid wave at the resonant layer induced by parametric instabilities. Computer simulation results of plasma heating from upper-hybrid mode conversion process were also reported by Lin et al. \cite{lin1982plasma}. Their study predicted that the energy of an obliquely incident microwave could be absorbed via wave breaking and the electron cyclotron heating induced by parametric instability. Theoretical study of the nonlinear effects concerning the three-wave decay interaction and modulations instabilities at the UH resonance layer was reported by Sharma et al. \cite{sharma1983nonlinear}. The excitation of upper-hybrid waves by a thermal parametric instability characterized by a four-wave interaction was reported by Lee et al. \cite{lee1983excitation}. 
 
 So far now, the UH resonance heating in the context of laser-plasma interaction has not been explored in detail. It is mainly because a very high magnetic field is required to magnetize the electrons in the time-scale of the incident laser frequency. This high requirement will be fulfilled in the near future soon as recent work by Nakamura et al. \cite{nakamura2018record} reported to have a record of 1200 Tesla indoor magnetic fields. Thus, it is just a matter of time the experimental study of UH resonance heating in the context of laser-plasma interaction would be feasible. However, there are some earlier works on the UH wave reported in the context of laser-plasma interaction. Upper-hybrid resonance absorption of laser radiation in a magnetized plasma in an inhomogeneous plasma was studied theoretically by Grebogi et al. \cite{grebogi1977upper}. Kitagawa et al. \cite{kitagawa1979upper} reported an experimental study on the upper-hybrid resonance absorption of $CO_2$-laser under a self-generated magnetic field in the plasma. Sodha et al. \cite{sodha1979excitation} reported the excitation of UH wave by a Gaussian EM beam, taking into account the nonlinear Ponderomotive force. Theoretical study on the wave-breaking phenomenon of relativistic upper-hybrid (UH) oscillations in a cold magnetoplasma was reported by Maity et al. \cite{maity2013wave}. They have also studied the breaking of linear and nonlinear electrostatic UH oscillation due to the phase mixing in the presence of an inhomogeneous external magnetic field \cite{maity2012breaking}. Computer simulation of upper hybrid and electron cyclotron resonance heating using a relativistic electromagnetic particle code was reported by Lin et al. \cite{lin1983computer}. Their study showed that the energy of an obliquely incident X-wave with a large angle of incidence converts into electrostatic Bernstein waves at the upper-hybrid resonance layer. In all the previous studies, detailed analysis and characterization of energy conversion processes from an X-wave propagating in an inhomogeneous external magnetic field were not explored thoroughly.
 
 In the present study, we have studied the propagation characteristics of an X-wave in a space varying external magnetic field ($B_0$) using PIC simulations. In our simulations, the laser was considered to be incident normal to the plasma surface, and intensity was chosen such that relativistic effects are absent in this study. It has been shown that as the laser enters inside the plasma, it follows the X-wave dispersion relation. The energy conversion processes from the X-wave to an electrostatic oscillation and eventually to the electron kinetic energy in the vicinity of the UH resonance layer ($\omega_l = \omega_{uh}$) have been investigated. In our study, we have also observed that at the resonance layer, electrostatic upper-hybrid oscillation breaks and converts its energy to the particle kinetic energy. Additionally, we have also observed that some parts of the electrostatic energy at the resonant layer convert to a low-frequency lower-hybrid mode and to the high-frequency harmonic radiations, which scatter in both directions from the resonance layer. These observations were not predicted in the earlier studies. Our study also reveals that the net energy conversion to the plasma particles suffers a loss by increasing the gradient scale lengths of the external magnetic field. This paper provides a comprehensive analysis of the outcome of upper hybrid oscillations in the presence of an inhomogeneous magnetic field.  
  
  This paper is organized as follows: in section \ref{picsim}, we have described the simulation set-up and provided the simulation parameters used in our study. Section \ref{obsrv} contains the observations obtained from the simulation. We have discussed different mode conversion processes from the incident laser beam in the various subsections. Finally, in section \ref{smry}, we provide a summary of our work.
 

\section{\it Simulation Details}
\label{picsim}

In this study, one-dimensional (1D) Particle-In-Cell (PIC) simulations have been carried out to study laser's interaction with plasma in the presence of an inhomogeneous external magnetic field. A fully relativistic, massively parallel PIC code, OSIRIS 4.0 \cite{Hemker,Fonseca2002,osiris} has been used for this purpose. The 1D simulation geometry considered here has a longitudinal extent of $2500 d_{e}$. Here, $d_e$ represents the electron skin depth $c/\omega_{pe}$, where c is the speed of light in vacuum and $\omega_{pe}$ defines the electron-plasma frequency corresponding to the equilibrium electron density $n_0$. Plasma boundary starts from $x = 100d_e$. The absorbing boundary conditions have been considered in both the directions for both fields and particles. The number of grid points for our simulation is considered to be $50000$, which corresponds to the grid size $dx = 0.05d_e$. The number of particles in each cell is considered to be $8$. We choose to normalize the time and length scales by $t_N=\omega_{pe}^{-1}$, and $x_N= c/\omega_{pe}= d_e$, respectively. The fields are normalized by $B_N=E_N= mc\omega_{pe}/e$, where $m$ and $e$ represent the mass and the magnitude of the charge of an electron, respectively. The external magnetic field is considered to be along $\hat z$ direction, and the plasma is along $\hat x$ direction, as shown in the schematic Fig. \ref{schmtc}(a).

 In our simulation, we have considered a laser pulse with a propagation vector $\vec{k}$ along $\hat x$ incident on the vacuum-plasma interface from the left side of the plasma. The electric field of the laser is considered to be in the X-mode configuration, i.e., oscillating along $\pm \hat y$, as shown in Fig. \ref{schmtc}(a). The longitudinal profile of the incident laser pulse is considered to be a polynomial function with a rise and fall time of $30\omega_{pe}^{-1}$, and it starts from $x = 80d_e$. The intensity of the incident laser pulse is considered to be approximately $3.04 \times 10^{13}$ $W cm^{-2}$, corresponding to the normalized vector potential $a_0=eE_l/m\omega_lc=0.05$. We have considered the dynamics of both electrons and ions in our study. The mass of the ions is considered to be approximately the same as the proton's mass, i.e., $M = 1840 m$. All the plasma and laser parameters that have been used in our simulations are provided in the table \ref{sim_tab} in normalized as well as standard units. The external magnetic field $B_0$ is considered such that it has a constant value equal to $B_0 = 2.5$ (in normalized unit) up to $x = 1000d_e$, and decreases linearly for $x>1000d_e$, as shown in the subplot (b) of Fig. \ref{schmtc}.

\begin{table}
\centering
\caption{Simulation parameters: In normalized units and possible values in standard units. }
\vspace{0.2cm}
	\begin{tabular}{|p{2.5cm}||p{2.5cm}||p{2.5cm}|}
		
		\hline
		\textcolor{black}{Parameters}& \textcolor{black}{Normalized Value}& 	\textcolor{black}{A possible value in standard unit}\\
		
		\hline
		\hline
		\multicolumn{3}{|c|}{\textcolor{black}{Laser Parameters}} \\
		\hline
		Frequency ($\omega_l$)&\centering$ 2.0 \omega_{pe}$& $1.78 \times 10^{14}$ rad/s\\
		
		\hline
		Wavelength&\centering$3.14c/\omega_{pe}$ &$10.6\mu m$\\
		\hline
		Intensity&\centering$a_{0} =0.05$ & $ 3.04 \times10^{13} W/cm^2$ \\
		\hline
		\multicolumn{3}{|c|}{\textcolor{black}{Plasma Parameters}} \\
		\hline
		Number density($n_0$)&\centering $1$ &$2.48 \times 10^{18}$ $ cm^{-3}$\\
		\hline
		Electron Plasma frequency ($\omega_{pe}$)& \centering$1$ &  $8.86\times10^{13}$rad/s\\
		\hline
		Electron skin depth ($c/\omega_{pe}$)& \centering$1$ & $3.37\mu m$ \\
		\hline
	\end{tabular}
		\label{sim_tab}

\end{table}

\section{\it Results and Discussion}
\label{obsrv}

\begin{figure*}[hbt!]
\centering
   \includegraphics[height = 5.5cm,width = 17.0cm]{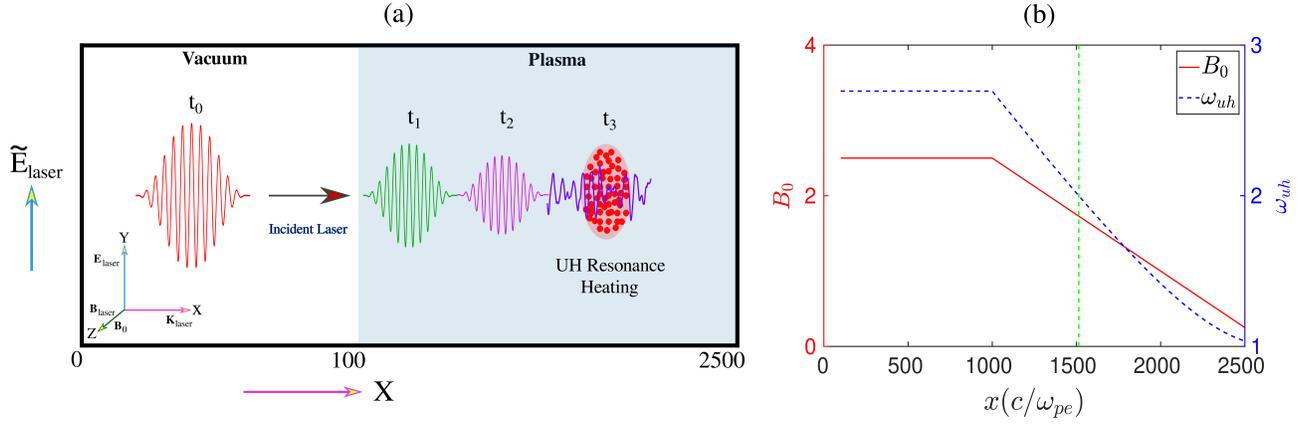}
   
   \caption{(a) The simulation setup and a summary of the physical processes observed in our study have been illustrated by this schematic. We have performed 1D PIC simulation with a laser pulse being incident from the left side of the simulation box on the vacuum-plasma interface at $x = 100$. The polarization of the incident laser was considered to be in the x-mode configuration, i.e., $\vec{E}_l \perp \vec{B}_0$. Here, $t_0$, $t_1$, $t_2$, and $t_3$ represent different times (in ascending order) of the simulation run. (b) The external magnetic field and corresponding upper hybrid frequency $\omega_{uh} = \sqrt{\omega_{pe}^2 + \omega_{ce}^2}$ have been shown as a function of $x$. Here, $\omega_{pe}$ and $\omega_{ce}$ represent electron plasma frequency and electron cyclotron frequency, respectively.}

  \label{schmtc}
\end{figure*}


It is well known that in a magnetized plasma, there are two types of waves that can propagate perpendicular to the external magnetic field ($\bf{B}_0$). One is the `ordinary' wave or O-wave where the electric field oscillates parallel to $\bf{B}_0$, and the other one is the `extraordinary' wave or X-wave where the electric field of the EM wave oscillates perpendicular to $\bf{B}_0$. The dispersion relation of high-frequency X-wave, where the ion motions can be neglected because of their large inertia, is given by \cite{goldston2020introduction},

\begin{equation}
\frac{c^2k^2}{\omega^2} = 1-\frac{\omega_{pe}^2\left(\omega^2-\omega_{pe}^2\right)}{\omega^2\left(\omega^2-\omega_{uh}^2\right)}
\label{disX}
\end{equation} 

Here, $\omega_{pe}$ and $\omega_{uh}$ represent electron plasma frequency and upper-hybrid frequency, respectively. It can be easily shown from equation \ref{disX} that high frequency X-wave has two distinct cutoff frequencies (defined as where $k \to 0$) as given by,

\begin{equation}
\omega_{R, L} = \frac{1}{2}\left[\pm \omega_{ce} + \left(\omega_{ce}^2 + 4\omega_{pe}^2\right)^{1/2}\right]
\label{cutf}
\end{equation}

Here, $+$ and $-$ signs stand for the `right-hand' and `left-hand' cut-off frequency, respectively and $\omega_{ce} = |eB_0/m|$ represents the electron gyration frequency. The dispersion relation given in equation \ref{disX} also shows that high frequency X-wave has a resonance ($k \to \infty$) at the frequency $\omega = \omega_{uh} = \sqrt{\omega_{pe}^2 + \omega_{ce}^2}$, known as the upper-hybrid resonance. The cut-offs and resonance define the pass and stop-bands for the propagation of EM wave inside the plasma. For a high frequency ($\omega$) X-wave, where the ion motion can be neglected, there exist two pass-bands: (i) $\omega_{uh}\geq \omega> \omega_{L}$, (ii) $\omega>\omega_R$, and one stop-band: $\omega_{R}> \omega> \omega_{uh}$.  

 In our PIC study, a laser pulse with a fixed value of frequency $\omega_l$ satisfying the condition $\omega_{uh}> \omega> \omega_{L}$ (i.e., lies in the pass-band) is sent into the plasma from the vacuum. The plasma starts from $x = 100d_e$ and extends up to $x = 2500d_e$. The electric field of the incident laser pulse is assumed to be in the X-mode configuration, as shown by the schematic in subplot (a) of Fig. \ref{schmtc}. The external magnetic field $B_0$ is considered to vary with space. The external magnetic field $B_0$ is chosen in such a way that it has a constant value ($B_0 = 2.5$ in normalized unit) up to $x = 1000d_e$ and then decreases linearly for $x>1000d_e$. The profiles of the external magnetic field and corresponding upper-hybrid frequency with the position $x$ have been shown in the subplot (b) of Fig. \ref{schmtc}. When the laser with frequency $\omega_l = 2.0\omega_{pe}$ incident on the magnetized plasma surface, it enters inside the bulk plasma as an X-wave and propagates through the plasma until it reaches the UH resonance point shown by the green dotted line in subplot (b) of Fig. \ref{schmtc}. At the UH resonance location, the electromagnetic energy of the X-wave is converted to electrostatic energy and essentially to electron kinetic energy. A summary of the observations of our study has been shown in the schematic in subplot (a) of Fig. \ref{schmtc}. We now present a detailed discussion of various features observed in our simulations in the following subsections.

\subsection{Propagation of X-wave in an inhomogeneous magnetic field}
\label{uh_osclltn}


\begin{figure*}[hbt!]
\centering
   \includegraphics[height = 7.5cm,width = 15.0cm]{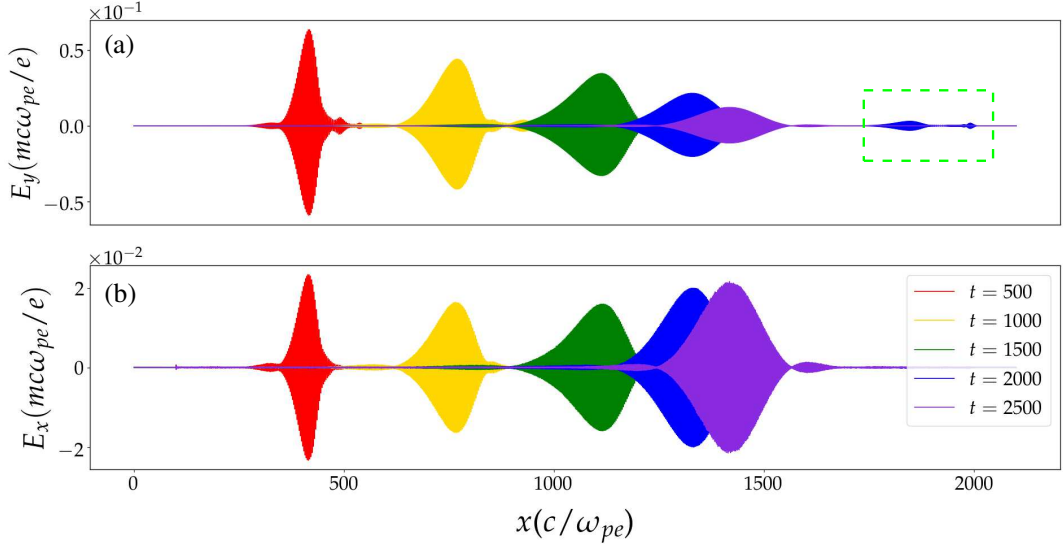}
   
   \caption{The $\hat y$ and $\hat x$ components of electric field $\widetilde{E}_y$ and $\widetilde{E}_x$ with respect to $x$ have been shown at different times of the simulation runs in subplots (a) and (b), respectively. Here, the $x$ locations of the fields at different times of the simulation run ($t = 500$ to $2500$) have been shown by the different colors. The structure highlighted by the green dotted box represent higher harmonics.}

  \label{field_in}
\end{figure*}

\begin{figure*}[hbt!]
\centering
   \includegraphics[height = 5.5cm,width = 14.0cm]{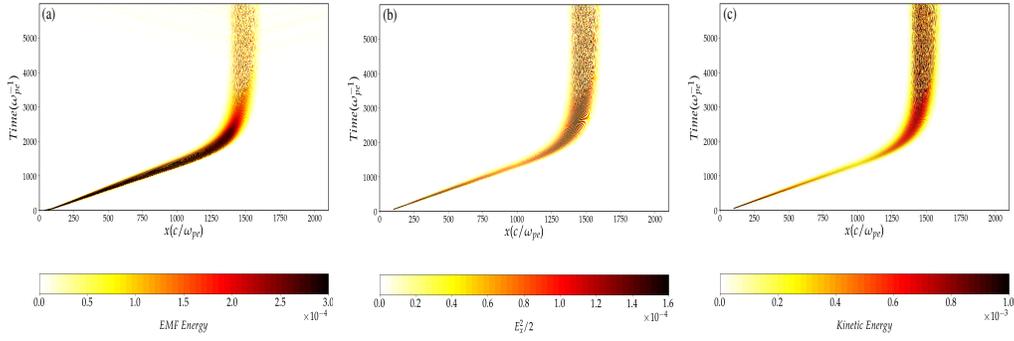}
   
   \caption{The variation of (a) electromagnetic field (EMF) energy, (b) energy associated with the longitudinal component of the electric field $E_x^2/2$, and (c) kinetic energy of electron with space and time have been illustrated in subplots (a), (b), and (c), respectively.}

  \label{eng_xt}
\end{figure*}


\begin{figure*}[hbt!]
\centering
   \includegraphics[height = 10.0cm,width = 14.0cm]{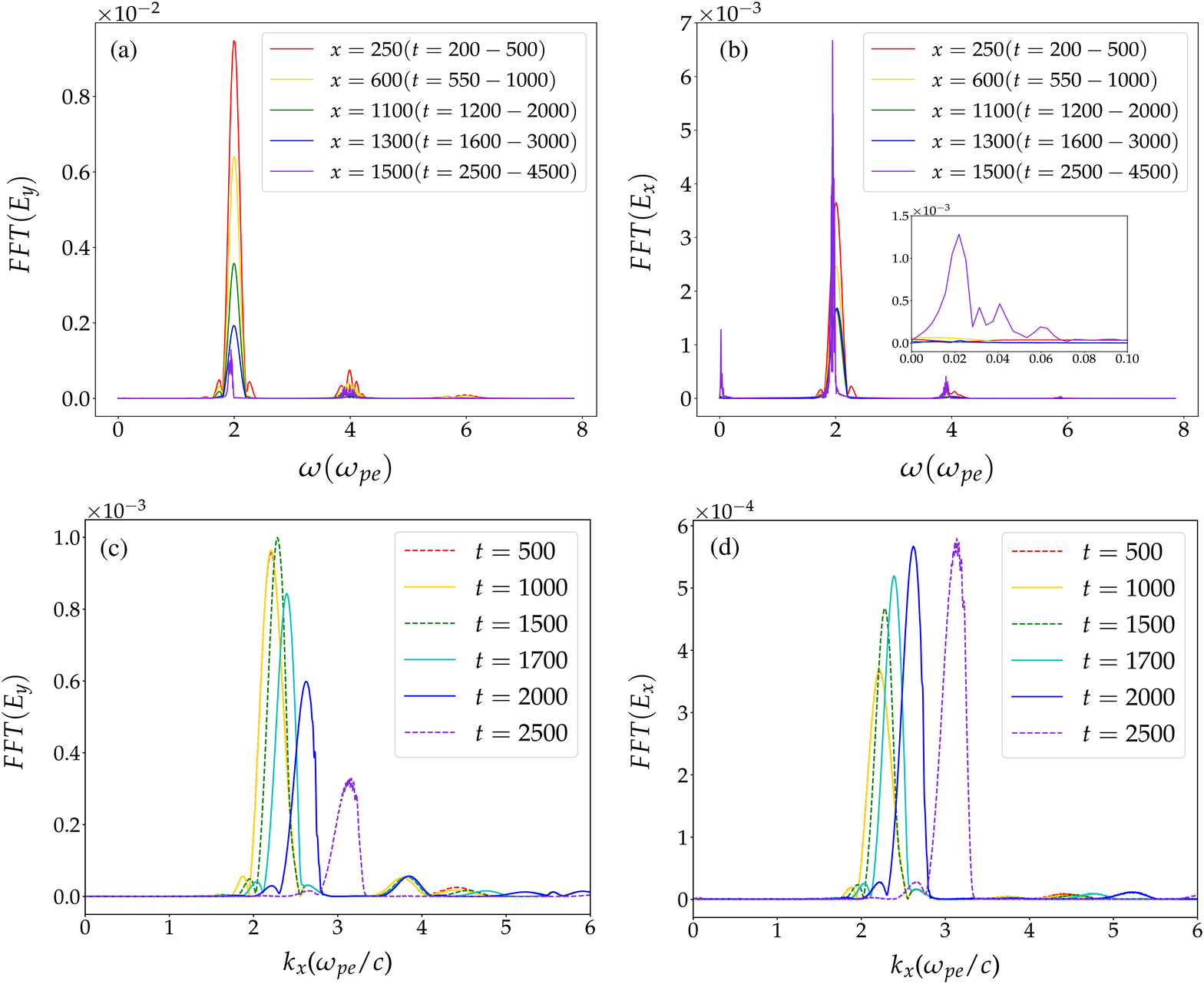}
   
   \caption{Time FFT (Fast Fourier Transform) of $\widetilde{E}_y$ and $\widetilde{E}_x$ at different $x$ locations have been shown in subplots (a) and (b), respectively. It is to be noticed that these FFT spectra have been evaluated at different time durations upto $t = 3500$. The space FFT spectra of $\widetilde{E}_y$ and $\widetilde{E}_x$ evaluated at different time of the simulation run have been elucidated by various colored line in subplots (c) and (d), respectively.}

  \label{fft_in}
\end{figure*}


The extraordinary-wave propagating perpendicular to $\vec{B}_0$ is partly electromagnetic and partly electrostatic. Thus, it has both transverse and longitudinal components of the electric field. At the upper-hybrid resonance point ($\omega_l = \omega_{uh}$), the wave will have a purely longitudinal component, but elsewhere it has a transverse component. For an X-wave, the electric field rotates in the clockwise direction, which is opposite to the electron rotation in the external magnetic field. Thus, electron-cyclotron resonance will not occur in the X-mode configuration. As the wave propagates through the plasma perpendicular to the inhomogeneous magnetic field $B_0$, the wave's frequency approaches towards the upper-hybrid frequency of the medium. Thus, the transverse component of electric field $E_y$ will decrease, but the strength of the longitudinal component $E_x$ will be increased. This has been shown in the subplots (a) and (b) of Fig. \ref{field_in}. The spatial distributions of the electromagnetic ($E_y$) and electrostatic ($E_x$) fields at different instants of time from $t = 50$ to $2500$ have been shown by different colored lines in subplots (a) and (b) of Fig. \ref{field_in}, respectively. It is seen that as the X-wave propagates through the plasma, the amplitude of $E_y$ decreases while the strength of the $E_x$ increases. It can be seen that there is a small segment of $E_y$ which has traveled even beyond the UH resonance layer. This has been highlighted by the green dotted box in subplot (a) of Fig. \ref{field_in}. The FFT spectrum, which is shown in Fig. \ref{fft_in} revealed that this structure is higher harmonic EM radiation generated from the X-wave as it travels through the magnetized plasma. Since the frequency of this harmonic radiation is higher than the `right-hand' cut-off frequency $\omega_R$, it passes through the UH resonance layer. The consequences of harmonic generation in the present context will also be discussed in section \ref{uh_heat}. 

 From the dispersion relation given in equation \ref{disX} it can be shown \cite{goldston2020introduction} that the phase velocity $v_p$ and the group velocity $v_g$ of a high frequency X-wave for $\omega \to \omega_{uh}$ can be approximately expressed as,

\begin{equation}
v_p = \frac{\omega}{k} \approx \frac{c\omega_{uh}\left( \omega_{uh}^2-\omega^2  \right)^{1/2}}{\omega_{ce}\omega_{pe}}
\label{eqvp}
\end{equation}

\begin{equation}
v_g = \frac{d\omega}{dk} \approx \frac{\left(\omega_{uh}^2-\omega^2\right)^{3/2}c}{\omega_{pe}\omega_{ce}\omega_{uh}}
\label{eqvg}
\end{equation}
 
It is seen from equations \ref{eqvp} and \ref{eqvg} that both the phase and group velocities of the X-wave decrease as we decrease the value of $\omega_{ce}$, i.e., with the decrease of external magnetic field $B_0$ and become zero for $\omega = \omega_{uh}$. This has been demonstrated in Fig. \ref{eng_xt} and also can be observed from Fig. \ref{field_in}. In Fig. \ref{eng_xt}, the distributions of electromagnetic field (EMF) energy, electrostatic energy $E_x^2/2$ associated with the longitudinal component of electric field $E_x$, and kinetic energy of electrons in space-time plane have been shown in subplots (a)-(c), respectively. It is seen that the profiles of energy start curving in the space-time plane at $x>1000d_e$ and become parallel to the time axis, accompanying a minimal spreading in $x$ at $x \approx 1505$. This can be understood by the fact that the group velocity of the wave decreases as it enters the decreasing regime ($x>1000d_e$) of the external magnetic field $B_0$. Finally, at $t\approx 2500$, as the wave-front touches the UH resonance layer ($x \approx 1505$), the wave-packet stops propagating further, and both its phase and group velocity become zero. At a later time, the energy of this wave-packet is converted to electrostatic energy and eventually dissipated to the kinetic energy of the plasma particles at the localized region surrounding the UH resonance layer, as shown in Fig. \ref{eng_xt}. This will also be discussed in further detail in the section to follow.

 The FFT (Fast Fourier Transform) spectra of the time series data of $E_y$ and $E_x$ in different time duration and at various $x$-locations have been shown in subplots (a) and (b) of Fig. \ref{fft_in}, respectively. As expected, it is clearly seen that the wave's frequency remains constant (same as the incident laser frequency $\omega_l = 2.0$) as it propagates through the medium. It is to be noticed that the amplitude of time FFT spectrum of $E_y$ evaluated at the location $x = 1500d_e$ and in-between time $t = 2500-4500$ becomes negligible compared to that of $E_x$. This has been clearly shown by the violet lines in the subplots (a) and (b) of Fig. \ref{fft_in}, which is a clear indication of the excitation of electrostatic UH oscillation with a frequency $\omega_{uh} = 2.0\omega_{pe}$. In the subplot (b), it is also seen that there is a distinct peak at very low frequency ($\omega \approx 0.023\omega_{pe}$) appeared in the FFT spectrum of $E_x$ evaluated in between $t = 2500 - 4500$ and at the location $x = 1500d_e$. This was not present in the FFT spectra of $E_x$ evaluated earlier, as shown in the inset of subplot (b), and also never appeared in the time FFT spectrum of $E_y$. The consequence of this will be discussed in the section to follow. Moreover, in each case, there are also peaks at $\omega = 4\omega_{pe}$ and $6\omega_{pe}$ in the FFT spectra, representing the higher harmonic radiations. The FFT spectra obtained from the distributions of $E_y$ and $E_x$ in space at different fixed values of time up to $t = 2500$ have been shown in subplots (c) and (d) of Fig. \ref{fft_in}, respectively. It is seen that the FFT spectra of both $E_y$ and $E_x$ show distinct peaks at a fixed value of $k_x \approx 2.23$ up to $t = 1000$. This is because the X-wave remains within the homogeneous regime of external magnetic field up to $t=1000$, as can be seen from Fig. \ref{field_in}. It is to be noticed that the wave's phase velocity ($\omega/k$) is less than the value of $c$. This is expected as the frequency of the X-wave is higher than the value of $\omega_{pe}$. Now, as we look at the space FFT spectra at later times, it is seen that the value of $k_x$ at which the FFT spectra show distinct peaks is shifted to the higher values. This is the consequence of the fact that the phase velocity decreases as the X-wave enters the decreasing regime of the external magnetic field. In the Fig. \ref{fft_in}, it is also seen that the peak value of FFT spectra of $E_y$ decreases while it increases for $E_x$. This is consistent with the results shown in Fig. \ref{field_in}.


\subsection{Electron heating and mode conversion at UH resonance layer}
\label{uh_heat}

\begin{figure*}[hbt!]
\centering
   \includegraphics[height = 7.5cm,width = 16.0cm]{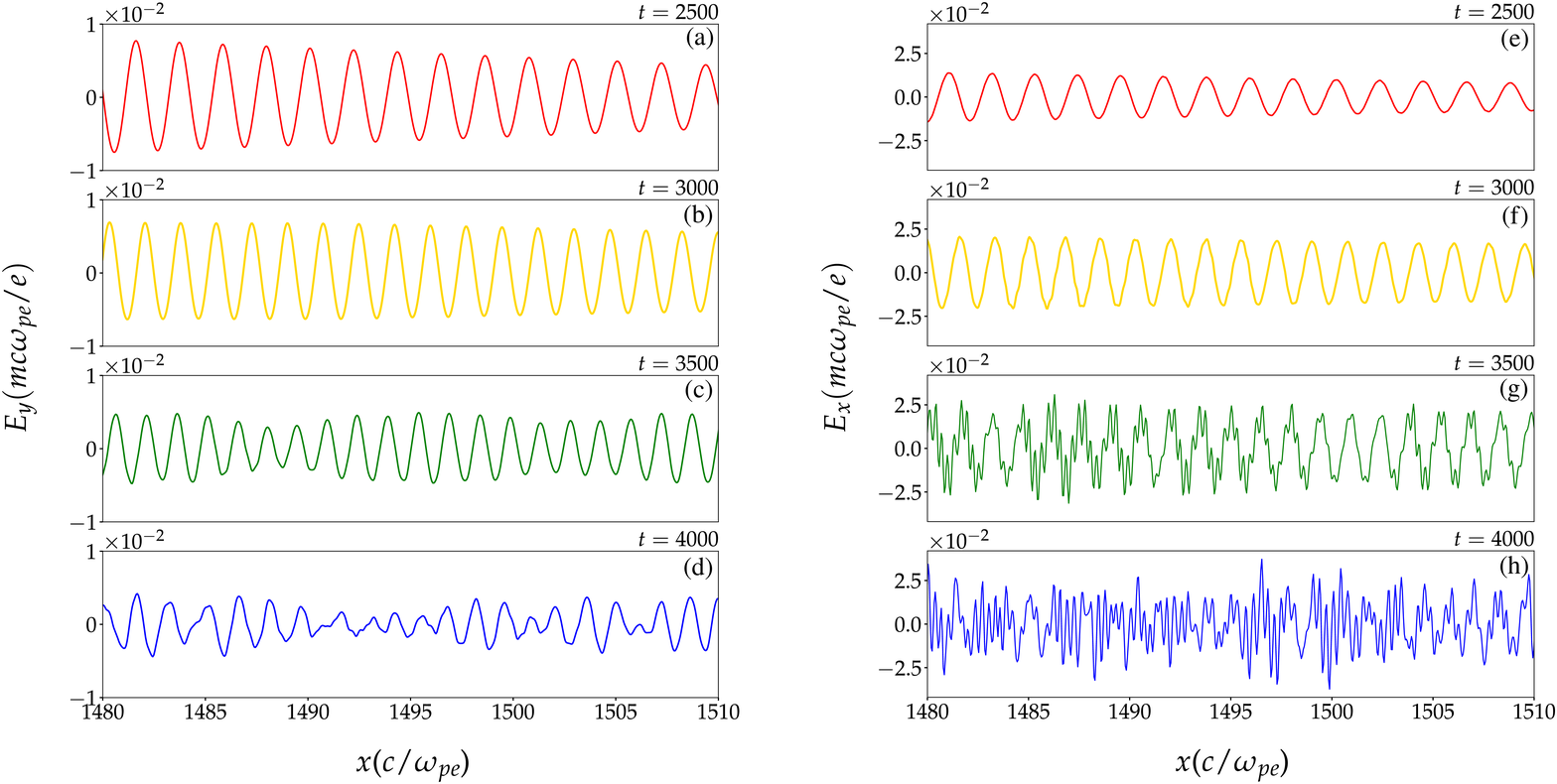}
   
   \caption{The zoomed view of $\hat y$ and $\hat x$ components of electric field $E_y$ and $E_x$ have been shown as function of $x$ for different instants of time in the later phase of the simulation run.}

  \label{field_zoom}
\end{figure*} 

\begin{figure*}[hbt!]
\centering
   \includegraphics[height = 7.5cm,width = 16.0cm]{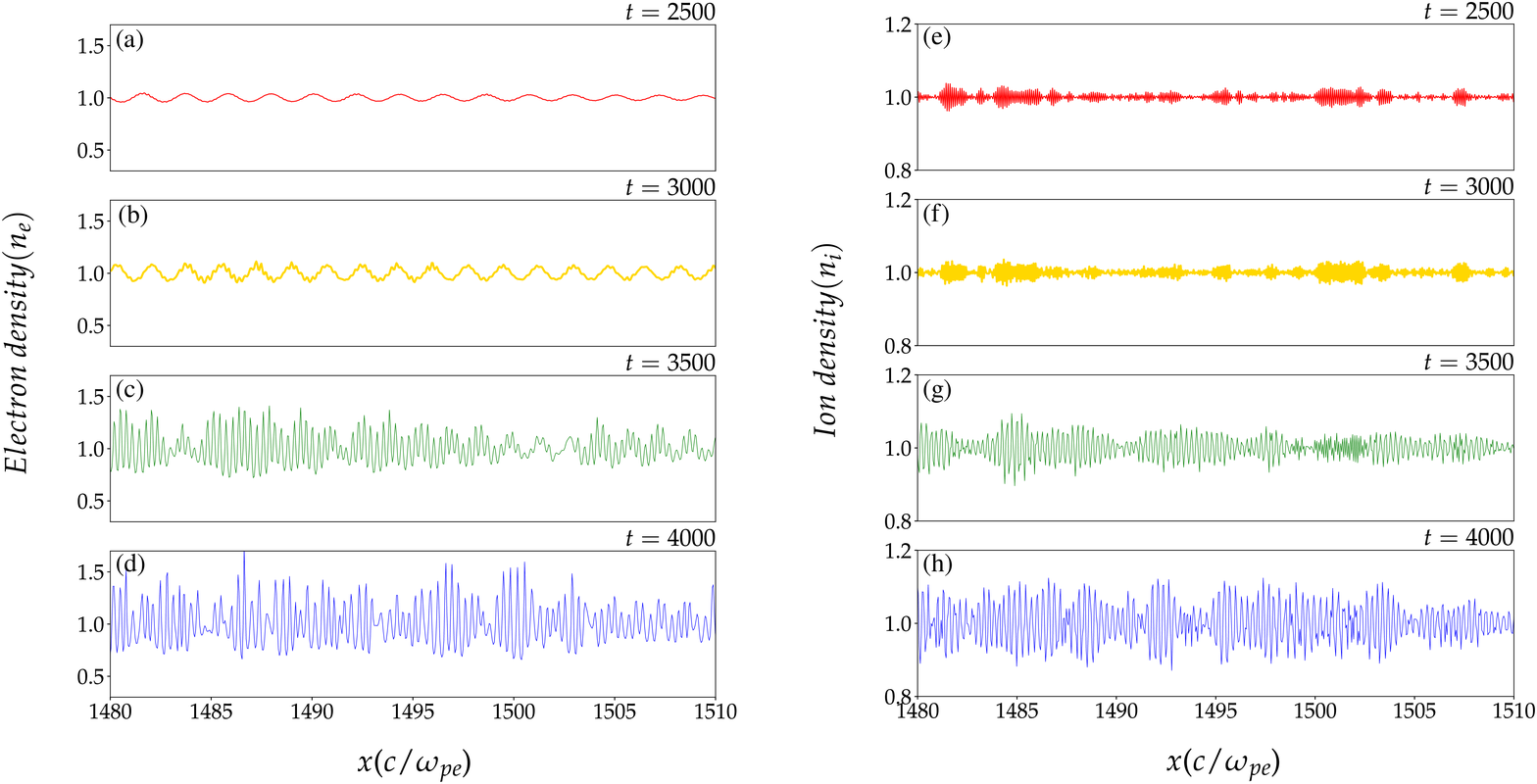}
   
   \caption{In subplots (a)-(d), the variation of electron density with $x$ have been shown at different instants of time. Subplots (e)-(h) illustrate the same for ion density.}

  \label{density_sub}
\end{figure*}
\begin{figure*}[hbt!]
\centering
   \includegraphics[height = 8.5cm,width = 12.0cm]{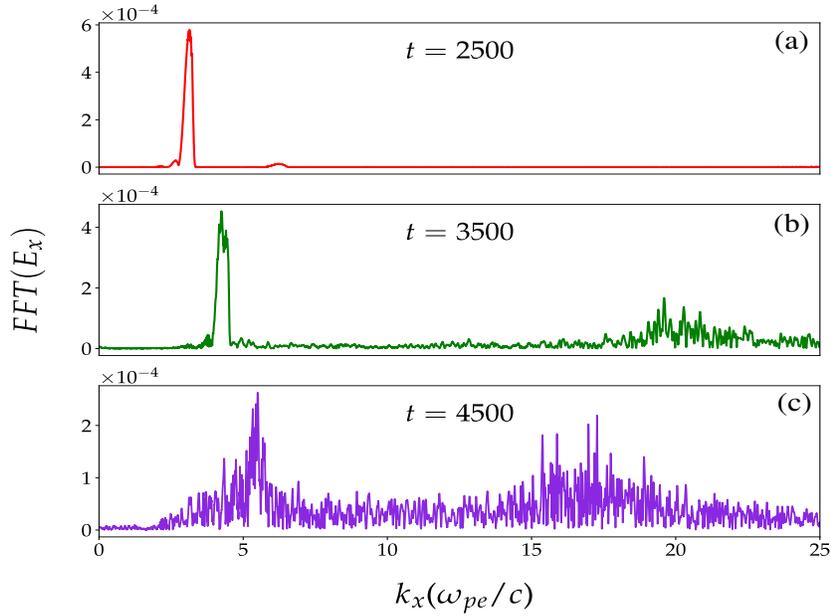}
   
   \caption{The space FFT spectra of longitudinal electric field $E_x$ at (a) $t = 2500$, (b) $t = 3500$, and (c) $t = 4500$ have been shown.}

  \label{fft_ex}
\end{figure*}

\begin{figure*}[hbt!]
\centering
   \includegraphics[height = 6.0cm,width = 16.0cm]{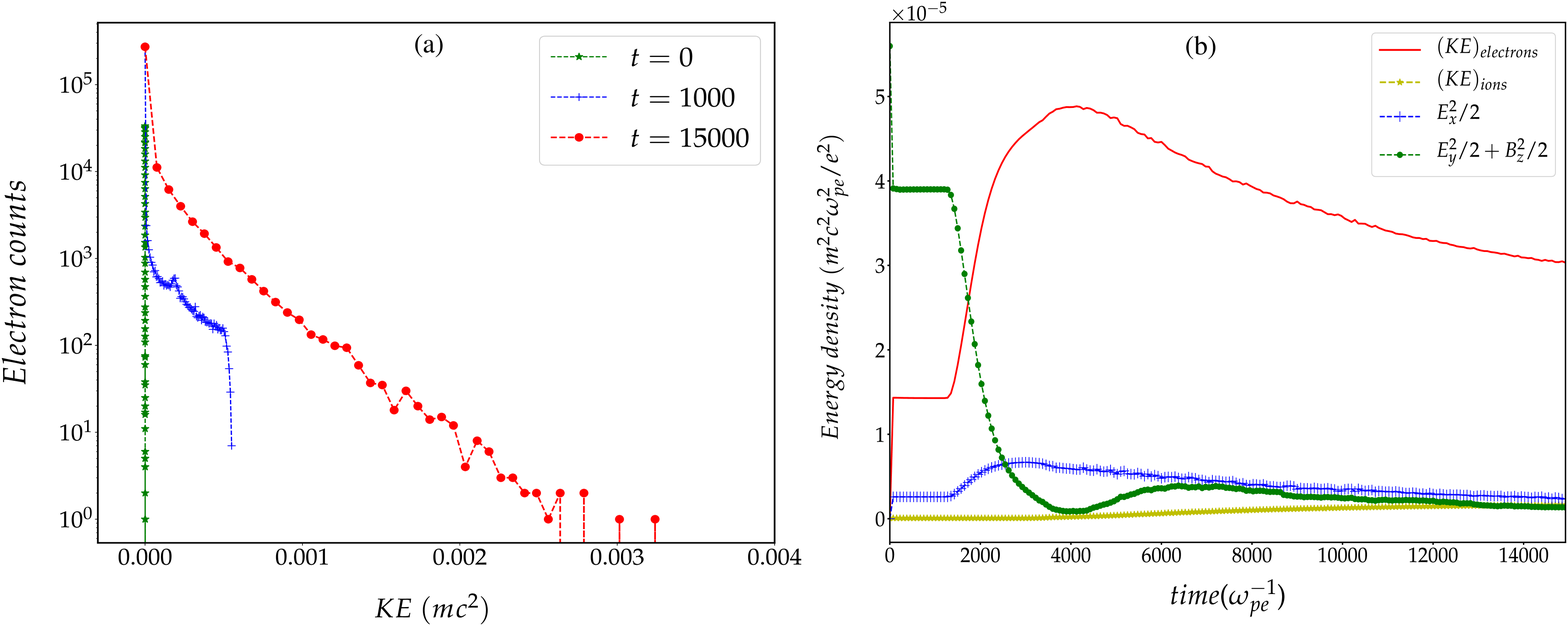}
   
   \caption{(a) Electron counts as a function of kinetic energy at three different instants of time of the simulation run have been shown. (b) The time variation of spatially averaged energy densities have been shown. Here, red, yellow, blue, and green colored lines represent kinetic energies of electrons and ions, energy density associated with the longitudinal electric field, and energy density associated with the transverse components of electromagnetic fields, respectively.}

  \label{eng_den}
\end{figure*}

\begin{figure*}[hbt!]
\centering
   \includegraphics[height = 7.5cm,width = 13.0cm]{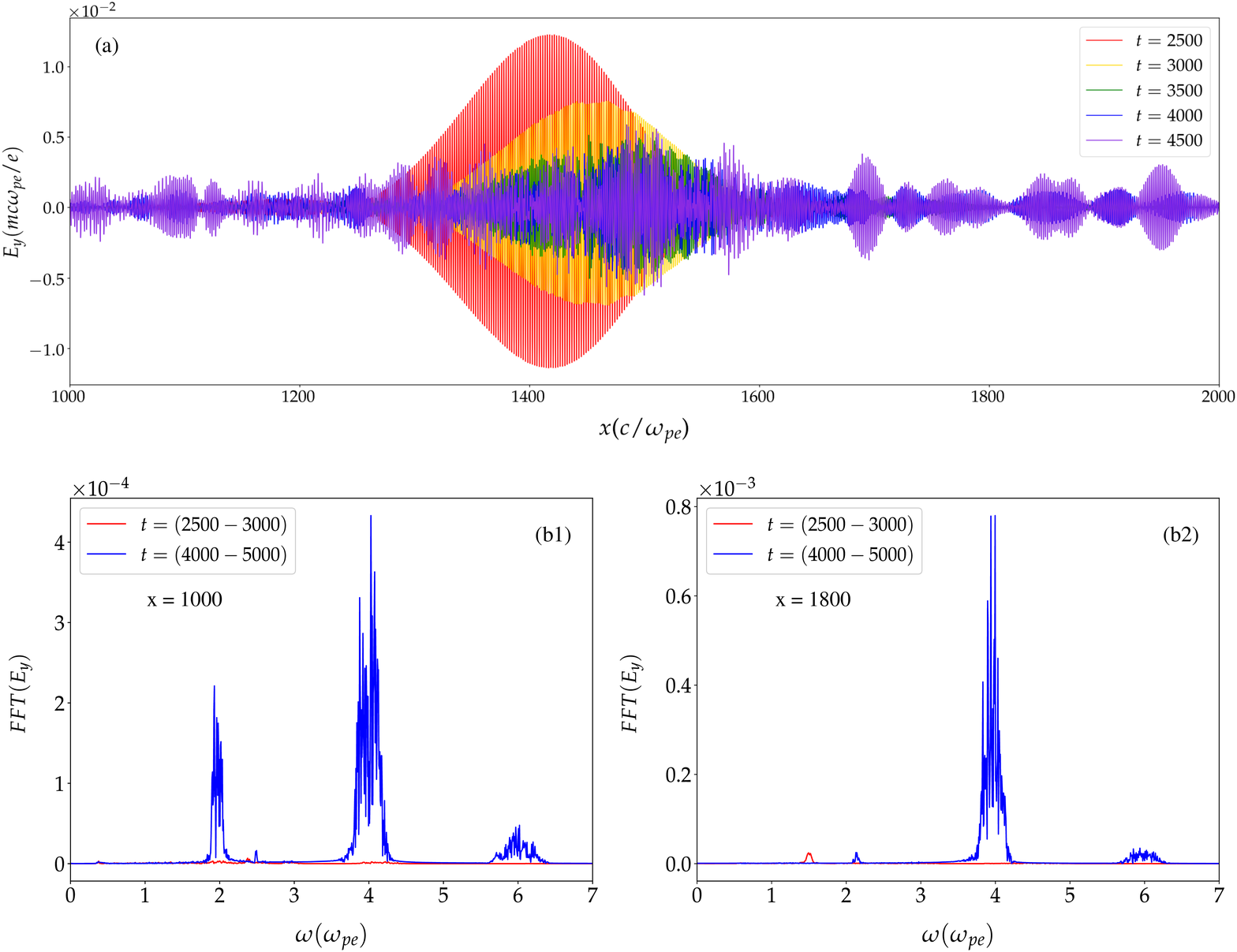}
   
   \caption{The space variation of $\hat y$ component of electric field has been shown at different times in the later stage of the simulation run in subplot (a). The time FFT spectra of $\hat y$ component of electric field $E_y$ evaluated at the locations $x = 1000$ and $x = 1800$ in the time durations $t = (2500-3000)$ (red), and $t = (4000-5000)$ (blue) have been shown in subplots (b1) and (b2), respectively.}

  \label{fld_fft_lt}
\end{figure*}


In this section, we will discuss the later stage of evolution of the X-wave in the vicinity of the UH resonance layer. To have a close look at the evolution of field profiles at UH resonance layer, we have shown in Fig. \ref{field_zoom} the space distributions of $E_y$ and $E_x$ surrounding the UH resonance layer at different fixed times from $t = 2500$ to $4000$. Amplitude of the transverse component of electric field $E_y$ increases up to $t = 3000$. This is because as soon as the font of the wave touches the resonance layer, its group velocity goes to zero, and wave-energy density continues to build up in the vicinity of the resonance layer. It is also to be noticed that up to $t = 3000$, amplitude of the electrostatic component of electric field $E_x$ also increases and both $E_x$ and $E_y$ have a nice sinusoidal form, as seen in subplots (a)-(b), and (e)-(f) of Fig. \ref{field_zoom}. As time goes on, the sinusoidal form of $E_y$ starts to deform, and wave-packets appear in the profile of $E_y$, as shown in subplots (c)-(d). At the same time, the amplitude of $E_x$ further increases, and its sinusoidal form starts to break, as shown in subplots (g)-(h) of Fig. \ref{field_zoom}. 

 We have also analyzed the time history of electron and ion density profile in space and shown in Fig. \ref{density_sub}. The electron density increases in vicinity of UH resonance layer whereas ions remain unperturbed (except noisy fluctuations) up to $t = 3000$, as shown in subplots (a)-(b), and (e)-(f) of Fig. \ref{density_sub}. It is also to be noticed that until this time, the electron density profile remains sinusoidal, which is consistent with the $E_x$ profile shown in subplots (e)-(f) of Fig. \ref{field_zoom}. As time evolves, the electron density further increases. Instead of having a sinusoidal profile, large-amplitude spikes appear in the density profiles, as shown in the subplots (c)-(d) of Fig. \ref{density_sub}. At the same time, disturbances in the ion density start to appear, and as time goes on, ion density fluctuations increase along with the appearance of density spikes. This has been shown in subplots (g)-(h) Fig. \ref{density_sub}. The ion density fluctuations are initiated in delayed time ($t \gtrsim 3000$) because the high frequency electrostatic UH wave needed to stay for a longer period in the vicinity of the UH resonance layer to perturb the ions having higher inertia. This is also apparent from the time FFT spectra of $E_x$ shown in the subplot (b) of Fig. \ref{fft_in}. It is seen that a distinct peak at the location $\omega \approx 0.0225 \omega_{pe}$ appears only at the later time ($t = 2500-4500$) in the FFT spectra of $E_x$, which is very close to the lower-hybrid frequency $\left(\omega_{lh} \simeq \sqrt{\omega_{pi}^2 + \omega_{ci}^2}\right)$. This result demonstrates mode conversion of the high-frequency upper-hybrid oscillation into a low-frequency lower-hybrid mode. The appearance of the spikes in the electron and ion density profiles confirms the breaking of electrostatic mode, which was also indicated in subplots (g)-(h) of Fig. \ref{field_zoom}. The breaking of UH oscillation in an inhomogeneous external magnetic field was reported by \cite{maity2012breaking}. In their study, the electrostatic UH oscillation breaks due to the phase mixing originated from the spatially varied external magnetic field. However, in our study, it is unclear whether the nonlinearity, inhomogeneity in $B_0$, or both, are responsible for breaking electrostatic UH waves.
 
 To further characterize the UH wave-breaking phenomena observed in our simulations, we have also evaluated the space FFT spectra of the electrostatic component $E_x$ in the later stage of the evolution. The space FFT spectra of $E_x$ evaluated at $t = 2500$, $3500$, and $4500$ have been shown in subplots (a)-(c) of Fig. \ref{fft_ex}, respectively. It is seen that only a sharp, distinct peak appears in the FFT of $E_x$ at $t = 2500$, and the particle value of $k_x$ where this peak is located can be exactly obtained from the theoretical dispersion relation of the X-wave. As time goes on, fluctuations at higher $k_x$ values also starts to appear in the FFT spectra of $E_x$, as shown in subplot (b) of Fig. \ref{fft_ex}. This is because after $t \approx 3500$, the electrostatic oscillation starts to break, as illustrated in Fig. \ref{field_zoom} and \ref{density_sub} and thus, energy flows to the higher and higher modes. As time increases further, it is seen from subplot (c) of Fig. \ref{fft_ex} that the power of the modes with higher values of $k_x$  increases. This clearly indicates particle heating, where energy flows from large to smaller scales.  
 
 To illustrate electron heating more distinctly, we have shown the electron energy distributions at three different instant of time of the simulation run in subplot (a) of Fig. \ref{eng_den}. It is seen that as time evolves, the distribution function generates high-energy tails indicating particle heating. In subplot (b) of Fig. \ref{eng_den}, we have shown the time evolution of space-averaged electron kinetic energy (red), ion kinetic energy (yellow), the energy associated with the electrostatic field, $E_x^2/2$ (blue), and the energy associated with the transverse component of the electric field, $E_y^2/2 + B_z^2/2$ (green dotted line). This figure clearly illustrates a complete picture of the energy-conversion process throughout the simulation run. As soon as the laser hits the plasma surface, electron kinetic energy increases. At the same time, electrostatic field energy ($E_x^2/2$), which was not present before, is also produced at the cost of electromagnetic energy ($E_y^2/2 + B_z^2/2$) of the incident laser pulse. Then, as long as the laser (X-wave) propagates inside the plasma within the homogeneous external magnetic field regime, electron kinetic energy, electrostatic energy, and electromagnetic energy remain constant. As the X-wave enters in the decreasing region of $B_0$ at $t \approx 1300$, the electrostatic energy and electron kinetic energy start to increase, whereas electromagnetic energy decreases. At the time $t \approx 2900$, when the X-wave has already reached the UH resonance layer, the electrostatic energy reaches a maximum value, whereas electron kinetic energy keeps increasing. Finally, at $t \approx 4000$, electron kinetic energy reaches a maximum value, and at the same time, the electromagnetic energy becomes almost zero. It is to be noticed that at this time, the electrostatic energy is in a decreasing trend from its peak value, as some parts of its energy are being converted to electron kinetic energy through wave breaking. However, it is interesting to notice that as time further increases, in between $t \approx 4000-6000$, the electromagnetic energy increases again from its minimum value. At the same time, both electrostatic and electron kinetic energy continues to drop until they get saturated. It indicates that there must be a reverse-conversion process where electrostatic energy gets converted to electromagnetic field energy. It is also to be noticed that after $t = 4000$, ions have also started to gain kinetic energy, as predicted in Fig. \ref{density_sub}.
 
 In order to have a deeper understanding of the mechanism involved in the mode conversion between electrostatic and electromagnetic field energies, we have observed the spatial distribution of $E_y$ at different instants of time after the X-wave reaches the UH resonance layer and shown in subplot (a) of Fig. \ref{fld_fft_lt}. It is interesting to see that at a later time ($t = 4500$), a part of $E_y$ is scattered in the form of wave packets along both sides of the resonance layer. It is to be noted that the original X-wave was neither supposed to cross the resonance layer nor reflect from the resonance layer. In order to further analyze this scattered electromagnet radiations, we have evaluated time FFT of $E_y$ at two different locations ($x = 1000d_e$ and $1800d_e$) in the both sides of the resonance layer and shown in the subplots (b1) and (b2) of Fig. \ref{fld_fft_lt}. The FFT spectra show that higher harmonics mainly dominate in these scattered radiations with frequencies $4.0\omega_{pe}$ and $6.0\omega_{pe}$. The higher harmonic generation in the presence of an external magnetic field in X and O-mode configurations were shown in detail in our previous study \cite{maity2021harmonic}. Let us try to understand briefly what happened at the UH resonance layer. As the X-wave reaches the resonance layer, its electromagnetic energy is eventually completely converted to electrostatic energy. A part of this electrostatic energy is then converted via wave breaking to the electron and ion kinetic energy resulting in heating. At the same time, some part of the electrostatic field ($E_x$) which is oscillating in the upper hybrid frequency ($2.0\omega_{pe}$), is getting converted to the high harmonic radiations in the presence of an external magnetic field and scattered in both sides of the resonance layer, which finally get absorbed at the boundaries. Thus, in this process, some parts of the electrostatic energy are also converted to electromagnetic energy, as seen in Fig. \ref{eng_den}.


\subsection{Effect of external magnetic field profile}
\label{B_prfl}
 
\begin{figure*}[hbt!]
\centering
   \includegraphics[height = 5.5cm,width = 14.5cm]{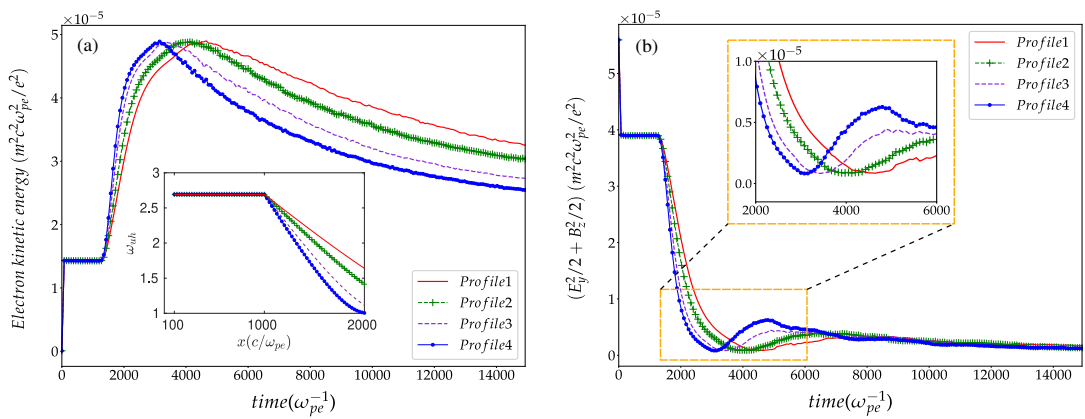}
   
   \caption{The variation of electron kinetic energy with time has been shown in subplot (a) for four different externally applied magnetic field ($B_0$) profiles: $Profile1$ (red solid line), $Profile2$ (green '+' marked line), $Profile3$ (violet dashed line), and $Profile4$ (blue dotted marked line). The space variation of upper hybrid frequency $\omega_{uh}$ corresponding to these four $B_0$ profiles has been shown in the inset of subplot (a). In the subplot (b), we have shown the time evolution of energy associated with the transverse component of electric field i.e., $E_y^2/2 + B_z^2/2$ for these profiles of $B_0$.}

  \label{ke_prfl}
\end{figure*} 
 
We have also studied the effect of the external magnetic field ($B_0$) profile on the energy conversion processes. For this purpose, we have considered four different $B_0$ profiles, and the corresponding upper-hybrid frequencies have been shown in the inset of Fig. \ref{ke_prfl}(a). It is to be noticed that in all four cases, $B_0$ is kept constant with a value 2.5 (in normalized unit) up to $x = 1000d_e$. We have shown the time evolution of electron kinetic energy in subplot (a) of Fig. \ref{ke_prfl} for all these four cases. It is seen that electron kinetic energy attains the maximum value earliest for the $B_0$ profile with the highest gradient. This is simply because for the steeper $B_0$ profile, the resonance layer is located at the smaller value of $x$, and thus, X-wave takes a shorter time to reach the resonance layer. It is also interesting to notice that the rate of decrease of the electron kinetic energy from its peak value is higher for the case with the $B_0$ profile having a steeper slope. Thus, the net energy converted irreversibly to the electrons, resulting in electron heating, will be high for the case with a $B_0$ profile having a less steep slope, as shown in the subplot (a) of Fig. \ref{ke_prfl}. This is because as the slope of the $B_0$ profile becomes steeper, the conversion efficiency of the electrostatic energy to the transverse component of electric field energy increases, as shown in the subplot (b) of Fig. \ref{ke_prfl}. Thus, the part of the electrostatic energy scattered away from the resonance layer in the form of harmonic radiation increases as we increase the inhomogeneity scale length of the external magnetic field. Let us now try to understand why the energy conversion process has such a dependency on the external magnetic profile. In our earlier work \cite{maity2021harmonic}, it has been shown that the efficiency of harmonic generation strongly depends on the value of an external magnetic field. The harmonic generation efficiency attains a maximum value for a particular value of $B_0$ where the condition $\omega_{ce} = 2\omega_l$ is satisfied and it increases with $B_0$ in the regime $\omega_{ce}<2\omega_l$. In the present study, at the upper-hybrid resonance layer ($\omega_l = \omega_{uh}$), the condition $\omega_{ce}<2\omega_l$ is always satisfied. We have shown that as the front of the X-wave touches the resonance layer, it stops propagating further. Since the X-wave has a finite longitudinal extend, it will face different values of $B_0$ at different x-locations in the vicinity of the resonance layer. However, the mean value of $B_0$ over the longitudinal length of the wave is higher for a $B_0$ profile with a steeper gradient. Thus, the harmonic generation efficiency increases for a steeper $B_0$ profile. As a result, more energy radiates away from the resonance layer via harmonics and is absorbed in the boundaries. Thus, the available energy to convert to electron kinetic energy becomes less.


\section{\it Summary}
\label{smry}

The characteristics of the extraordinary wave (X-wave) originated from the interaction of laser beam with a magnetized plasma is studied using PIC simulations. It has been shown that the group velocity and phase velocity of X-wave changes as it propagates under an inhomogeneous magnetic field and essentially goes to zero as it reaches the upper-hybrid resonance point. The energy conversion from the X-wave to the UH electrostatic mode at the resonance layer has been shown. This electrostatic wave essentially breaks and converts its energy to the electrons in the vicinity of the resonance layer. Additionally, our study also reveals two additional processes involved in the decaying of electrostatic oscillations. A part of the electrostatic energy converts to the lower-hybrid mode causing ion density fluctuations and heating. A significant portion of the energy associated with the electrostatic mode also converts to the high-frequency harmonic EM radiation and is scattered away from the resonant location. In our study, we have shown that the net energy absorbed by the electrons depends on the profile of the external magnetic field. Our study reveals that electrons gain more kinetic energy for an external magnetic profile with a gentler slope. On the other hand, the conversion efficiency to the harmonic radiations in the vicinity of the resonance layer can be increased by increasing the inhomogeneity scale length of the external magnetic profile.

\section{\it Acknowledgment} 

The authors would like to acknowledge the OSIRIS Consortium, consisting of UCLA and IST (Lisbon, Portugal) for providing access to the OSIRIS 4.0 framework which is the work supported by NSF ACI-1339893. This research work has been supported by the 
 J. C. Bose fellowship grant of AD (JCB/2017/000055) and
the CRG/2018/000624 grant of DST. The authors thank IIT Delhi HPC facility for computational resources.


\bibliography{UH_ref}





\end{document}